\newcommand{\C}{$^{12}$C}

\newcommand{\N}{$^{14}$N}

\documentclass[nolinenumbers]{aastex631}

\shorttitle{Shock-Induced Burning in SNe Ia}
\shortauthors{Brady \& Zingale}
\graphicspath{{./}{figures/}}

\begin{document}

\title{Numerical Treatment of Shock-Induced Nuclear Burning in Double Detonation Type Ia Supernovae}


\author[0009-0003-1338-6336]{Ryan Brady}
\affiliation{Dept. of Physics and Astronomy, Stony Brook University, Stony Brook, NY 11794-3800, USA}

\author[0000-0001-8401-030X]{Michael Zingale}
\affiliation{Dept. of Physics and Astronomy, Stony Brook University, Stony Brook, NY 11794-3800, USA}

\begin{abstract}
We present a benchmark problem to assess the treatment of shock-induced nuclear burning in the context of double detonation Type Ia supernovae. In a stratified white dwarf model, we implement a shock-detection criterion that suppresses burning in zones characterized by compression and significant pressure gradients, controlled by a tunable parameter, $f_{\text{shock}}$. One-dimensional simulations, using the open-source \texttt{Castro} suite, were conducted across three treatments—burning fully enabled, and burning suppressed with $f_{\text{shock}} = {2}/{3}$ and $f_{\text{shock}} = 1$—across three spatial resolutions (5.0, 2.5, and 0.3125 km). At the finest resolution, the burning-enabled and $f_{\text{shock}} = 1$ models converge, while the $f_{\text{shock}} = {2}/{3}$ front continues to show slight offset behavior.  Since most simulations are carried out at much lower resolutions, our tests support the idea that burning in shocks should always be disabled in practice.  We also
observe that the behavior of lower-resolution simulations remains extremely sensitive to the choice of $f_{\text{shock}}$.
\end{abstract}

\keywords{Hydrodynamical simulations(767) --- Type Ia supernovae(1728) --- Open source software(1866) --- GPU computing(1969)}

\section{Introduction} \label{sec:intro}

Recent observations of peculiar Type Ia Supernovae (SNe Ia)—such as those with unusual luminosities or spectra \citep{liu:2022, gonzalez:2023}—suggest that a portion may originate from sub-Chandrasekhar mass progenitors. A leading model for these is the double detonation mechanism \citep{nomoto:1982}, where a detonation occurs in an accreted helium-rich envelope of a white dwarf (WD), likely triggered by compressional heating from a companion star whose Roche lobe overflows. This detonation drives an outward shock and an inward compression wave, which may ignite carbon fusion at the WD center, resulting in a secondary detonation and a full thermonuclear explosion akin to a standard SN Ia.

Computational studies \citep{gronow:2020, boos:2021, rivas:2022} have examined the conditions for double detonations using varied approaches to reaction networks, hydrodynamics, and numerical treatments. While broadly consistent, these models still face uncertainties—particularly regarding burning within shock regions. This issue was first postulated astrophysically by \citet{fryxell:1989}, who argued that numerical diffusion smears shocks across zones, artificially initiating burning. They proposed disabling burning in regions with strong pressure jumps and compressive flow. \citet{papatheodore:2014} extended this to cellular detonations in uniform media, reinforcing the practice. As a result, most double detonation simulations now suppress burning in shocks. In this note, we revisit the issue with a test problem that utilizes a stratified WD model, assessing convergence and providing a more realistic benchmark for stellar detonation models.

We use \texttt{Castro} \citep{castro,castro_joss}, an open-source simulation code that solves multicomponent compressible hydrodynamic equations. Reactions are coupled to the hydrodynamics using the simplified spectral deferred corrections (SDC) method from \citet{zingale:2022}. For our work, we use the criterion:
\begin{equation}
\frac{\left| (\nabla P - \rho \textbf{g}) \cdot \textbf{U}_{\text{cell}} \right|}{P_{\text{cell}} |\textbf{U}_{\text{cell}}|} > f_{\text{shock}} \quad \quad \nabla \cdot \textbf{U} < 0
\end{equation}
to tag a zone as containing a shock. We treat the velocity vector of the zone, $\textbf{U}_{\text{cell}}$, as a unit vector and project the pressure gradient, $\nabla P$, in its direction. The negative velocity divergence requirement ensures we tag shocks in regions of compression. In one dimension, this is analogous to the method in \citet{fryxell:1989} and adopted in other works. We also consider that hydrostatic pressure in a stratified medium, given by density $\rho$ and gravitational acceleration $\textbf{g}$, cannot drive a shock. Lastly, the arbitrary detection threshold $f_{\text{shock}}$ is of $\mathcal{O}(1)$.

\section{Results} \label{sec:simulations}

\begin{figure}[b]
    \centering
    \gridline{
        \fig{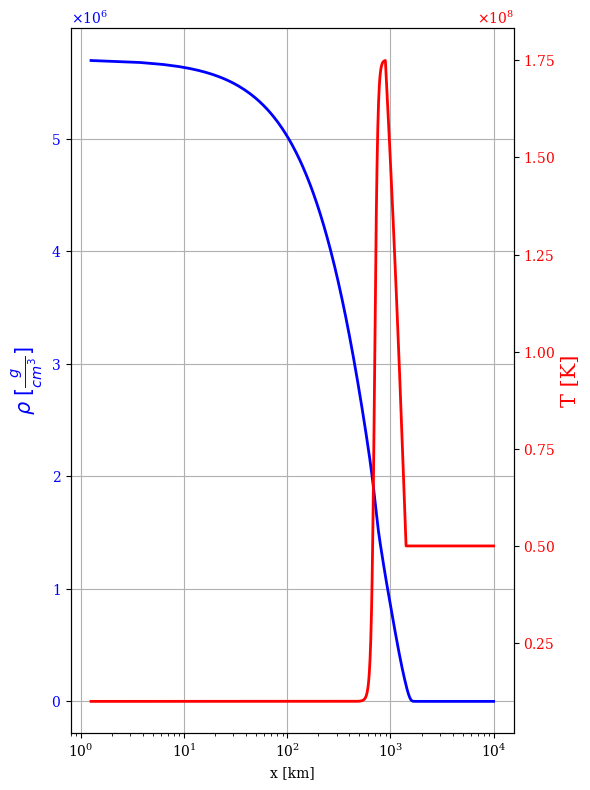}{0.45\textwidth}{}
        \fig{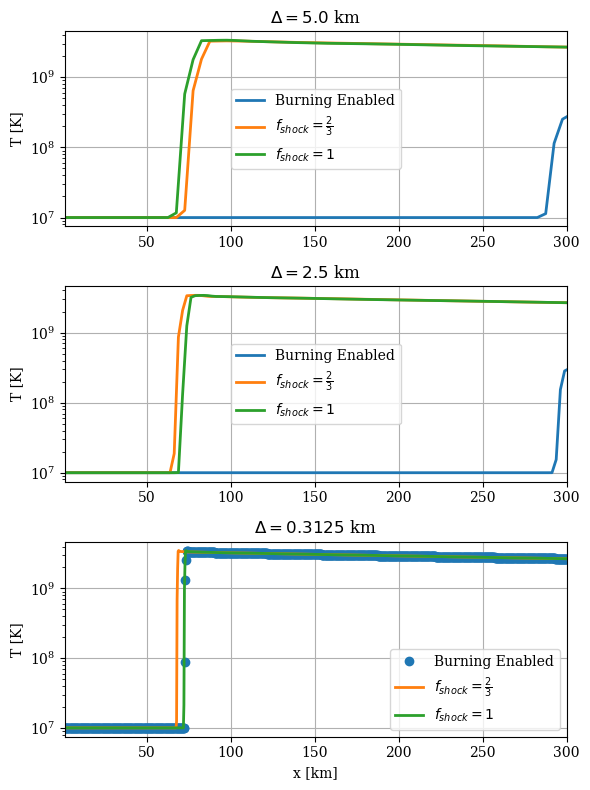}{0.45\textwidth}{}
    }
    \caption{\label{fig:results} (\textit{Left}) Initial hydrostatic profile without the temperature perturbation $T_{\mathrm{pert}} =1.14 \times 10^9$ K. \textit{(Right)} Snapshot of our shock probe at $t = 0.08$ s across spatial resolutions of 5.0 km (top), 2.5 km (middle), and 0.3125 km (bottom).  Since the burning-enabled profile
    overlaps the $f_\mathrm{shock} = 1$ profile at the finest resolution, we highlight it with data markers.}
\end{figure}

One-dimensional simulations were performed in a plane-parallel geometry that mimics conditions near the WD--accretion layer interface in a double detonation SN Ia progenitor. The model is based on a $1.1\,\text{M}_\odot$ WD with a $0.05\,\text{M}_\odot$ helium-rich layer from \citet{zingale:2024}. The WD core is composed of 0.5 $^{12}$C and 0.5 $^{16}$O at an isothermal temperature of $T_\mathrm{core} = 10^7\,\mathrm{K}$. A hyperbolic tangent transition region of width $\delta = 50\,\mathrm{km}$ connects to the accreted layer, with $T_\mathrm{base} = 1.75 \times 10^8\,\mathrm{K}$ and a composition of 0.99 $^4$He and 0.01 $^{14}$N. An isoentropic drop brings the system to $T_\mathrm{lo} = 7.5 \times 10^7\,\mathrm{K}$, after which the domain remains isothermal. The simulation domain spans 4000\,km, with symmetry at the core (left) boundary and outflow at the top (right). A constant gravitational acceleration of $\textbf{g} = -1.1742 \times 10^9\,\mathrm{cm\,s^{-2}}$ is applied throughout. A temperature perturbation of $T_\mathrm{pert} = 1.14 \times 10^9\,\mathrm{K}$ at the base of the envelope initiates the detonation. Simulations used the 22 nuclei and 94 reaction rate \texttt{subch\_simple} nuclear network described in \citet{chen:2023}.

The right panel of Figure~\ref{fig:results} shows the temperature evolution at $t = 0.08$\,s for the three spatial resolutions and shock burning treatments. At the coarsest resolution ($\Delta = 5.0$\,km), only the model with shock burning enabled does not propagate directly into the WD core. This agrees with \citet{zingale:2024}, who found successful double detonations when burning in shocks were not suppressed at this resolution; had they been, the flame may have directly ignited the WD core, violating the double detonation model by bypassing the compression wave induced secondary ignition. At intermediate resolution ($\Delta = 2.5$\,km), the relative behavior of the shock treatments changes. While $f_{\text{shock}} = 2/3$ lagged behind $f_{\text{shock}} = 1$ at low resolution, it now leads slightly. None of the treatments have converged at this scale, indicating resolution sensitivity. At the highest resolution ($\Delta = 0.3125$\,km), the temperature profiles for $f_{\text{shock}} = 1$ and the burning-enabled case converge, suggesting that shock burning suppression becomes unnecessary when the shock structure is well-resolved. However, the $f_{\text{shock}} = 2/3$ front still leads, implying that intermediate thresholds can influence detonation behavior even at very fine resolution.

While a sufficiently resolved shock does not require the treatment of prohibiting burning, these results confirm that it is required at coarser resolutions to ensure that a significant amount of fuel will not burn inside of the numerically broadened shock, consistent with the results of \citet{fryxell:1989} and \citet{papatheodore:2014}. The choice of $f_{\text{shock}}$ thus represents a trade-off that depends sensitively on resolution: lower-resolution simulations benefit from suppression (e.g., $f_{\text{shock}} = 2/3$), while higher-resolution models converge toward an $f_{\text{shock}} =1$ solution with full shock burning enabled. 
We also note that the mutlidimensional situation is more complex, as explored briefly in \citet{zingale:2024}.

\begin{acknowledgments}

\texttt{Castro} is freely available at \href{https://github.com/AMReX-Astro/Castro}{https://github.com/AMReX-Astro/Castro} and \citet{castro:2025}. The metadata, input files, and data analysis notebook for this work can be found in \citet{brady:2025}. This work was supported by DOE/Office of Nuclear Physics grant DE-FG02-87ER40317, by New York Space Grant Consortium award 80NSSC20M0096, and by Undergraduate Research and Creative Activies (URECA) at Stony Brook University.

\end{acknowledgments}

%

\vspace{5mm}
\facilities{ORNL Frontier}


\bibliography{refs}{} 
\bibliographystyle{aasjournal}



\end{document}